\documentclass[final,1p,times,number]{elsarticle}

\usepackage{lineno,hyperref}
\usepackage{tikz}
\usetikzlibrary{arrows.meta,calc,decorations.markings,math,arrows.meta,shapes.arrows, fadings}
\usepackage{physics}
\usepackage{tikz-feynman, amsmath}
\modulolinenumbers[5]
\graphicspath{{img/}}
\usepackage{comment}
\usepackage{float}
\usepackage{units}
\usepackage{amssymb}
\usepackage{xcolor}

\usepackage{subcaption}
\usepackage{caption}

\definecolor{darkblue}{rgb}{0.1,0.2,0.6}

\journal{Annals of Physics Special Issue: Localisation 2020}

\begin{document}

\begin{frontmatter}

\title{Many-body localization in large systems: Matrix-product-state approach}

%\begin{comment}

\author{Elmer V.~H.~Doggen}
\address{{Institute for Quantum Materials and Technologies, Karlsruhe Institute of Technology, 76021 Karlsruhe, Germany}}
\address{{Institut f\"ur Theorie der Kondensierten Materie, Karlsruhe Institute of Technology, 76128 Karlsruhe, Germany}}
\author{Igor~V.~Gornyi}
\address{{Institute for Quantum Materials and Technologies, Karlsruhe Institute of Technology, 76021 Karlsruhe, Germany}}
\address{{Institut f\"ur Theorie der Kondensierten Materie, Karlsruhe Institute of Technology, 76128 Karlsruhe, Germany}}
\address{Ioffe Institute, 194021 St. Petersburg, Russia}
\author{\mbox{Alexander~D.~Mirlin}}
\address{{Institute for Quantum Materials and Technologies, Karlsruhe Institute of Technology, 76021 Karlsruhe, Germany}}
\address{{Institut f\"ur Theorie der Kondensierten Materie, Karlsruhe Institute of Technology, 76128 Karlsruhe, Germany}}
\address{L.~D. Landau Institute for Theoretical Physics RAS, 119334 Moscow, Russia}
\address{Petersburg Nuclear Physics Institute, 188300 St.~Petersburg, Russia}
\author{Dmitry G.~Polyakov}
\address{{Institute for Quantum Materials and Technologies, Karlsruhe Institute of Technology, 76021 Karlsruhe, Germany}}

\begin{abstract}

Recent developments in matrix-product-state (MPS) investigations of many-body localization (MBL) are reviewed, with a discussion of benefits and limitations of the method. This approach allows one to explore the physics around the MBL transition in systems much larger than those accessible to exact diagonalization.  System sizes and length scales that can be controllably accessed by the MPS approach are comparable to those studied in state-of-the-art experiments. Results for 1D, quasi-1D, and 2D random systems, as well as 1D quasi-periodic systems are presented.
On time scales explored (up to $t \approx 300$ in units set by the hopping amplitude), a slow, subdiffusive transport in a rather broad disorder range on the ergodic side of the MBL transition is found.
For 1D random spin chains, which serve as a ``standard model'' of the MBL transition, the MPS study demonstrates a substantial drift of the critical point $W_c(L)$ with the system size $L$:  while for $L \approx 20$  we find $W_c \approx 4$, as also given by exact diagonalization, the MPS results for $L = 50$--100 provide evidence that the critical disorder saturates, in the large-$L$ limit, at $W_c \approx 5.5$.  For quasi-periodic systems, these finite-size effects are much weaker, which suggests that they can be largely attributed to rare events.  For quasi-1D ($d\times L$, with $d \ll L$) and 2D ($L\times L$) random systems, the MPS data demonstrate an unbounded growth of $W_c$ in the limit of large $d$ and $L$, in agreement with analytical predictions based on the rare-event avalanche theory.

\end{abstract}


\end{frontmatter}

\section{Introduction}

Philip W.~Anderson considered what happens to quantum particles when they move through a disordered medium \cite{Anderson1958a}. He showed that disorder can fully halt transport. According to the celebrated scaling theory of localization \cite{Abrahams1979a}, even weak disorder is sufficient to prevent transport in the thermodynamic limit in one-dimensional (1D) and two-dimensional (2D) geometries (more accurately, this applies to all Wigner-Dyson symmetry classes in 1D and to two of them---orthogonal and unitary---in 2D). For three-dimensional (3D) systems, a transition from the phase of localized to the phase of delocalized states occurs, which can be driven by the strength of disorder or another parameter of the system \cite{Kramer1993a, Evers2008a}. Such localization has been observed in various disordered media, ranging from photonic lattices \cite{Lahini2008a} and optical fibers \cite{Karbasi2012a} to dilute Bose-Einstein condensates \cite{Billy2008a, Roati2008a}.

While the Anderson-localization problem is highly non-trivial and very rich by itself, the complexity of the problem is further strongly increased when the interaction between particles is included. The added complexity has several interrelated facets. One of them concerns the zero-temperature quantum phase transitions (and corresponding phase diagrams) in an interacting disordered system. We will not touch this (also very rich) field in the present brief review which focuses on the physics of highly excited states (i.e., those at non-zero energy density). The natural expectation in such a situation is that interactions between particles disturb interference effects that are responsible for Anderson localization, thus leading to delocalization. Indeed, Fermi's golden rule would suggest a non-zero quasiparticle width (i.e., a finite decay rate) at any non-zero temperature. However, with sufficiently strong disorder, the application of Fermi's golden rule (that assumes a continuous spectrum of final states) may be invalidated, leading to a breakdown of the above expectation and to the emergence of a localized phase of the many-body interacting system at non-zero temperature. This was in fact pointed out by Anderson himself, together with Fleishman, in the seminal paper \cite{Fleishman1980a}. The field of many-body localization (MBL), as it is known today, has its roots in theoretical works from the mid-2000s examining this scenario in detail \cite{Gornyi2005a, Basko2006a}. Experiments on MBL-type systems started in the 2010s, especially using ultracold atoms \cite{Lucioni2011a, Schreiber2015a}, where disorder is implemented using laser fields.

With the rapid development of numerical algorithms and increasing numerical resources, the problem of MBL has become within the reach of computational approaches, initially in one dimension. Oganesyan and Huse \cite{Oganesyan2007} and \v{Z}nidari\v{c} \emph{et al.}~\cite{Znidaric2008a} pioneered the numerical study of the MBL problem using two complementary approaches. One of them \cite{Oganesyan2007}, exact diagonalization, fully solves the many-body problem through computational brute force. This approach has no approximations, but since the computational complexity of a many-body system increases exponentially with the system (e.g., $\sim 2^L$ for a spin-$\frac{1}{2}$ chain of length $L$), the method is limited to small systems (up to $L \approx 20$). Luitz \emph{et al.}~\cite{Luitz2015a} used a particularly numerically efficient implementation of exact diagonalization and found evidence for an MBL transition---i.e., a transition from the thermalizing (ergodic) to the localized phase. At the same time, the apparent critical behaviour provided by the corresponding scaling analysis turned out to be inconsistent with the Harris criterion, which is an indication of the fact that the system sizes reached by exact diagonalization ($L \approx 20$) are too small for investigation of the critical properties.

The second approach, which was used in Ref.~\cite{Znidaric2008a} to demonstrate MBL at strong disorder, belongs to a class of numerical methods based on \emph{matrix product states} (MPS) \cite{Schollwock2011a}. An MPS is a type of tensor network that encodes a many-body wave function in terms of a variational ansatz. A key feature of the MPS approach is that the accuracy of the method can be controlled through the \emph{bond dimension} $\chi$, which can be tuned to interpolate between a product state, where there is no entanglement, and a maximally entangled state (i.e., it allows for a description of the wave function without approximations). Not surprisingly, the computational efficiency of describing maximally entangled states using MPS is not better than that of exact diagonalization, so that the MPS approach is only useful for systems that are relatively weakly entangled. In practice, one often starts in an (unentangled) product state, and computes the dynamics until the entanglement grows out of control. Thus, the approach is usually limited to finite times because of the growth of entanglement. We note that MPS approaches can also be used to probe the static properties, especially those of ground states that are often weakly entangled, but here we focus primarily on dynamics.

The rate of the entanglement growth is therefore of significant importance for the applicability of MPS approach. In the MBL phase, for strong disorder, analytical and numerical studies find a slow, logarithmic growth of entanglement \cite{Bardarson2012a}. This is very beneficial for the investigation of the MBL phase by the MPS method. On the other hand, in the opposite limit of weak disorder, the system quickly thermalizes and shows ergodic behaviour \cite{Luitz2015a}, in agreement with the eigenstate thermalization hypothesis (ETH) \cite{Polkovnikov2011a}. Of much physical interest is the behaviour of the system in a numerically broad range of intermediate disorder strengths, i.e, around  the MBL transition, with the ergodic side of the MBL transition turning out to be particularly intriguing. What is the behaviour of the entanglement in this regime? The answer to this question is of paramount importance for the applicability of the MPS framework to the study of the MBL problem around the transition point.

Another key question concerns the role of the \emph{type} of disorder in the system. Numerous experiments deal with interacting \emph{quasiperiodic} (rather than truly random) systems. While such systems have no innate randomness, experimentally they also show an MBL transition \cite{Schreiber2015a}. The distinction between the two is crucial from a theoretical perspective because the physics of random systems is expected to be strongly affected by rare regions, whereas such regions do not occur in quasiperiodic systems. In particular, the type of disorder is expected to qualitatively influence the \emph{type} of dynamics, in the sense that a subdiffusive dynamics is expected \cite{Luitz2017a} on the basis of Griffiths effects. Yet, the experiment indicates subdiffusive behaviour also for quasiperiodic systems.

One more question concerns the case of higher-dimensional systems (i.e., beyond 1D), where the effects of rare, weakly disordered regions should become {\rm stronger}, which was predicted \cite{DeRoeck2017a, Thiery2017a, Gopalakrishnan2020a} to destabilize the MBL phase in the thermodynamic limit at any fixed strength of disorder $W$. Yet, also in this case cold-atom experiments show evidence for an MBL transition at finite disorder \cite{Choi2016a}. Importantly, however, the experiments explore finite systems, with the number of atoms being of the order of $100$, while the thermodynamic limit in the rare-region analysis is reached very slowly. In order to verify the predictions of the rare-region theory and to compare to the experiment, it is thus of crucial importance to perform computations in a broad range of system sizes, at least up to those studied experimentally. This goal is by far out of reach of exact diagonalization, which again demonstrates an important advantage of the MPS approach.

These questions have been addressed in a series of papers by the present authors \cite{Doggen2018a, Doggen2019a, Doggen2020a}, where it was demonstrated that the MPS approach is very useful for the investigation of the vicinity of the MBL transition, including its ergodic side, in large systems and in a variety of settings. The key results, along with related advances by other researchers, are reviewed in this article.  For more general reviews of MBL, the reader is referred to Refs.~\cite{Nandkishore2015a, Altman2015a, Abanin2017a, Alet2018a}.

\section{Analytical background}
\label{sec:avalanche}

\subsection{Approaches based on the structure of the perturbative expansion}
\label{sec:pert}

Analytical approaches to MBL-type problems were pioneered by Fleishman and Anderson in  Ref.~\cite{Fleishman1980a}, where it was argued that short-range interactions in a system with localized single-particle states do not necessarily destroy localization at finite temperature $T$. The argument was based on the perturbative analysis of the single-particle decay rate caused by the excitation of localized particle-hole pairs. Later, higher-order electron-electron scattering processes were addressed in terms of an Anderson-localization problem on a certain graph in Fock space in Ref.~\cite{Altshuler1997a}, where the broadening of hot single-particle states in a quantum dot (at zero temperature) was studied.

More recently, Refs.~\cite{Gornyi2005a, Basko2006a} have studied the effect of a weak short-ranged interaction in disordered systems where all states are localized in the absence of the interaction (in particular, in 1D geometry). These works posed the following question: How does the system evolves from a metal (at high temperature $T$) to an insulator (at low $T$), assuming no coupling to an external bath? The scaling of the ratio of a higher-order matrix element to the corresponding many-body level spacing with the order of the perturbation theory was obtained by means of an approximate mapping on the Cayley-tree in Fock space in Ref.~\cite{Gornyi2005a} and a self-consistent resummation of the perturbation series in Ref.~\cite{Basko2006a}. Both these works concluded that there is a finite-$T$ localization-delocalization transition manifested in the temperature dependence of the quasiparticle decay rate, and thus of the conductivity, which is zero for $T<T_c$ and finite for higher $T$. The critical temperature $T_c$ of the MBL transition was found to be
\begin{equation}
\label{Tc}
T_c \sim \frac{\delta_\xi}{\alpha\ln(1/\alpha)}\,,
\end{equation}
where $\delta_\xi$ is the single-particle level spacing within the localization volume and $\alpha\ll 1$ is the dimensionless interaction strength. Up to the logarithmic factor, Eq.~\eqref{Tc} can be obtained by equating the typical interaction matrix element to the typical level spacing of three-particle states for the decay of a single-electron excitation, which provides a relation to the argument in Ref.~\cite{Fleishman1980a}.  The logarithmic factor in Eq. (\ref{Tc}) is analogous to (and has the same origin as) the one  in the problem of localization on a Bethe lattice with a large connectivity.  The scaling (\ref{Tc}) was rederived by the analysis of the perturbation series within the forward-scattering approximation in Ref.~\cite{Ros2015a}. A formal proof of the existence of the MBL phase in a 1D model, which used a formulation of stability of  the perturbation series in terms of local integrals of motion as well as certain additional assumptions, was provided in Ref.~\cite{Imbrie2016a}.

In Ref.~\cite{Gornyi2017a} the effect of spectral diffusion driven by the diagonal interaction matrix elements (discarded in Refs.~\cite{Gornyi2005a, Basko2006a}) was taken into account. The analysis of the corresponding higher-order matrix elements resulted in a different scaling of the transition temperature for $\alpha\ll 1$:
\begin{equation}
\label{Tc12}
T_c \sim \frac{\delta_\xi }{\alpha^{1/2}\ln^\mu(1/\alpha)},
\end{equation}
with $0\leq \mu\leq 1/2$.
The essence of the effect is that the interaction-induced resonant transitions between quasiparticle states in a many-body system shift single-particle energy levels, which enhances the occurrence of new resonances, thus favoring delocalization. As a result, it turns out to be sufficient to find a single resonance in a certain spatial region to trigger further transitions, instead of requiring each single-particle state to have a resonant three-particle partner. Consequently, for $\alpha\ll 1$, delocalization persists down to lower temperatures than that given by Eq. (\ref{Tc}). We note, however, that the difference between Eqs.~(\ref{Tc}) and (\ref{Tc12}) is not essential for sufficiently strong interaction, $\alpha\sim 1$, which is used in most of the numerical simulations. When translated into the language of the XXZ spin chain at infinite temperature---the paradigmatic model used for numerical studies of MBL~\cite{Znidaric2008a}, see Eq.~(\ref{eq:xxzham}) below---both the MBL thresholds (\ref{Tc}) and (\ref{Tc12}) become the conditions on the disorder strength, $W=\mathcal{O}(1)$ for $\Delta\sim J\sim 1$.

\subsection{Random regular graphs}

A closely related direction of the analytical studies of MBL, inspired by the ideas of  Ref.~\cite{Altshuler1997a}, is based on the connection of MBL to Anderson localization on tree-like structures such as random regular graphs  (RRG) and their relatives. An RRG is a finite graph with randomly connected vertices (hence ``random''), but with the same branching number---the number of legs at each vertex (hence ``regular''). Locally, an RRG has the structure of a Cayley tree, but, in contrast to it, all sites of the RRG have the same connectivity, i.e., it has no boundary. The Anderson model on such a graph can be viewed as a toy model of MBL. In this analogy, RRG vertices play the role of basis many-body states (i.e., eigenstates of the non-interacting Hamiltonian), while the hopping between them the interaction matrix elements. This model thus conforms to  the situation when the energies of the basis many-body states are uncorrelated. The true many-body problem can be exactly represented by a model on a certain graph in the Fock space with tree-like properties but with non-trivial correlations. The effect of spectral diffusion partly reshuffles energies of many-body states connected by interaction matrix elements, thus rendering the structure closer to (although still different from) that of the RRG model.  Despite the difference, the RRG model captures many key properties of the genuine many-body problem. For a more detailed review of the RRG model and its relations to the MBL problem, we refer the reader to Ref.~\cite{tikhonov2021_RRG}.

An important advantage of the RRG model is that it can be studied analytically in a fully controllable way by means of a field-theoretical approach and the saddle-point analysis justified for large system size \cite{Tikhonov2019a,tikhonov19critical}. The resulting analytical predictions, including, in particular, the ergodicity of the delocalized phase, the localized character of the critical point, and critical scaling at the transition, have been supported by numerical simulations \cite{Tikhonov2019a,tikhonov19critical,Tikhonov2016a,garcia-mata17,metz2017level,biroli2018,PhysRevResearch.2.012020,tikhonov2020eigenstate}.

The fact that the RRG model can be solved analytically is very useful since it can be used to test performance of exact diagonalization. The maximum Hilbert-space size of the RRG system that can be studied realistically via exact diagonalization is $\sim 10^6$ vertices, which corresponds to a spin chain of the length $\approx 20$. Comparing exact-diagonalization results with analytical predictions in the limit of infinite system size, one finds rather strong finite-size effects for the critical disorder and the scaling at criticality. In particular, the apparent critical point drifts substantially towards stronger disorder with increasing system size.
As we describe in this review, a very similar trend is observed in genuine interacting systems. The analysis of RRG saddle-point equations in Ref.~\cite{tikhonov19critical} by population dynamics has allowed one to reach effectively the system with Hilbert-space size of nearly $10^{20}$, which corresponds to spin chains of length larger than $L=60$. It was found that such system sizes are indeed needed to reach a good convergence of the numerically found critical behaviour to analytical predictions. In true many-body systems, finite-size effects are expected to be still stronger than on RRG, due to rare-region effects. We thus come again to the conclusion that numerical studies of systems with $ \approx 100$ sites are of great need for computational investigation of the MBL physics.

\subsection{Rare regions and avalanches}

The perturbative analysis of MBL focuses on typical disorder configurations, thus neglecting Griffiths-type effects associated with exponentially rare events. Rare regions with locally anomalously strong or anomalously weak disorder may potentially affect  the localization and transport properties in large systems, where
the probability of finding at least one such rare region becomes substantial. In particular, strong-disorder spots on the ergodic side of the MBL transition were argued
to slow down the dynamics in the system, leading to subdiffusive transport \cite{Agarwal2015a, Vosk2015a, Potter2015a, Agarwal2016a}. As another manifestation of a potential impact of rare regions, the MBL phase can be destabilized through the growth of ``ergodic spots,'' which naturally arise in the system in regions with anomalously weak disorder. Such ergodic spots can grow over time, by absorbing the neighboring localized regions---the process that has been termed an \emph{avalanche}  \cite{DeRoeck2017a, Thiery2017a, Gopalakrishnan2020a}. The many-body delocalization point is then the point where this growth becomes unbounded, so that the whole system eventually thermalizes. The avalanche description has specific implications for the qualitative behaviour of MBL-type systems, and on the way this behaviour changes for various geometries and types of disorder, as we are going to detail.

The avalanche picture assumes the presence of a rare region with anomalously weak disorder, such that, in the absence of couplings to the rest of the system, this ``seed" would be ergodic. The exact many-body states in this ergodic (thermal) seed are assumed to be described by random matrix theory, while the rest of the system, excluding the ergodic seed, is considered to be many-body localized. The Hamiltonian of the initially localized part of the system is then described in terms of local integrals of motion (LIOM)---``spins,'' characterized by the localization length $\xi$ that depends on the stength of disorder. Switching on the coupling between the subsystems, the ratio of the coupling matrix element for hybridizing a spin located at distance $r$ from the spot to the \textit{many-body} level spacing inside the ergodic spot  should satisfy
\begin{equation}
e^{-r/\xi} N_s^{-1/2}/N_s^{-1}\gtrsim 1.
\label{ratio}
\end{equation}
Here, $N_s$ is the dimension of the space of many-body states inside the spot and the factor $N_s^{-1/2}$ in the matrix element accounts for the ergodic (random-matrix) scaling of the local operators. In a 1D spin chain, $N_s=2^\ell$ for an ergodic spot of size $\ell$ (in units of the lattice constant). It follows then that, once
\begin{equation}
\xi > 2/\ln 2,
\end{equation}
an emergent avalanche in the chain cannot be stopped \cite{DeRoeck2017a, Thiery2017a}. This condition translates into the condition on the disorder strength $W$, which does not depend on the strength of the interaction, in contrast to the conditions of the type (\ref{Tc}) and (\ref{Tc12}).

The probability of finding a rare thermal seed of the required size depends on the system dimensions and dimensionality: the larger the system, the higher this probability, and hence delocalization is more probable than in smaller systems. Thus, in numerical studies of MBL, the estimated position $W_c$ of the MBL transition is expected to drift towards stronger disorder with increasing system size. At variance with 1D systems, this growth of $W_c$ with the system size as predicted by the avalanche theory turns out to be unbounded in systems of higher dimensionality. This is because the size $N_s$ of the Hilbert space associated with the ergodic seed increases, then, with increasing its linear dimension faster than an exponential \cite{DeRoeck2017a}. In particular, avalanches in 2D systems of size $L \times L$
give rise to the following dependence of the critical disorder strength $W_c(L,L)$ on $L$ ~\cite{Gopalakrishnan2019a}:
\begin{equation}
\ln W_c(L,L)\sim \ln^{1/3}L^2.
\label{eq:2dWc}
\end{equation}
Further, the avalanche-induced scaling of the critical disorder strength $W_c(L,d)$ in a quasi-one-dimensional strip of width $d \gg 1$ (in units of the lattice constant) and length $L \ge d$ is found to be \cite{Doggen2020a}
\begin{equation}
W_c(L,d) \sim \begin{cases} \exp[c_1 \ln^{1/3}(Ld)], &\quad d \le L<L_*(d),\\
2^d, &\quad L>L_*(d).
\end{cases}
\label{eq:q1dWc}
\end{equation}
with an estimate $c_1 \approx 1.57$ for the numerical coefficient $c_1$. Here the crossover length $L_*(d)$ is given by
\begin{equation}
L_*(d) \sim d^{-1} \exp \left[  \left( \frac{d \ln 2}{c_1} \right)^3 \right] \,.
\label{eq:L-star}
\end{equation}
We see that in the 1D limit ($d = 1$), Eq.~\eqref{eq:q1dWc} yields a finite critical value in the thermodynamic limit $W_c(\infty,1) \sim 1$, as has been already stated above. In the quasi-1D regime, i.e., for increasing $L \gg d$ at fixed $d$ and, we also  have a finite value of $W_c(\infty,d)$, which, however, increases exponentially with $d$. Finally, in the 2D regime, $d \sim L$,  Eq.~\eqref{eq:q1dWc} takes the form of Eq.~\eqref{eq:2dWc}, predicting an unbounded growth of $W_c$ with $L$. Thus, according to the avalanche theory, the MBL gets destabilized in the 2D case by rare events in the thermodynamic limit if the disorder strength $W$ is kept fixed. At the same time, for large but finite $L$, there is a well-defined MBL transition at an $L$-dependent critical disorder $W_c(L)$.

It is worth emphasizing that the avalanche theory has a rather phenomenological character. Its conclusions are crucially dependent on the validity of the central assumption that the growing seed retains its ``most ergodic'' character, i.e., that it can be modeled by random matrix theory. While numerical support for this assumption has been obtained in simplified models, it is of crucial importance to compare the predictions of the avalanche theory to numerical simulations of genuine MBL systems of sufficiently large size---which points to the importance of MPS-based computational investigations of the MBL transition beyond 1D. Below, we will overview numerical results obtained by the MPS approach that do provide support for the avalanche theory.

Utilizing the ideas of the strong-randomness renormalization-group (RG) framework, several phenomenological RG theories have been proposed to describe the MBL transition. Recent works have expanded on the idea of the competition of localizing and delocalizing rare-event effects, suggesting that the MBL transition is of the Berezinskii-Kosterlitz-Thouless (BKT) type  \cite{Dumitrescu2019a, Goremykina2019a, Morningstar2019a, Morningstar2020a}. A key prediction from this class of theories is that the critical point is itself localized (similar to the critical point in RRG models) and, therefore, only weakly entangled. In this brief review, we discuss evidence for such a scenario from the viewpoint of the MPS approach. The localized character of the critical point is highly beneficial for the investigation of the critical region and of a part of the ergodic phase (with $W$ comparable to $W_c$), because of a relatively slow growth of the entanglement.

\section{MBL problem and matrix product states}

\subsection{Matrix product states}

Here we briefly describe the MPS concept and several key MPS algorithms that have been used to describe the MBL transition. The reader is referred to the extensive general reviews on the density matrix renormalization group (DMRG) \cite{Schollwock2011a}, 1D and 2D tensor networks \cite{Cirac2020a}, and the computation of dynamics using MPS \cite{Paeckel2019a} for more in-depth details.

An MPS describes the many-body wave function in the following way. For simplicity, consider a 1D system on a lattice of length $L$ with site index $i = 1,\ldots,L$, with a single, binary degree of freedom $\sigma_i$ on each site. This could be a spin-$1/2$ chain, or a particle degree of freedom restricted to single occupancy, i.e., spinless fermions or hard-core bosons. The wave function can then be written as:
\begin{equation}
|\Psi \rangle = \sum_\sigma c_{\sigma_1 \ldots \sigma_L} |\sigma_1, \ldots, \sigma_L \rangle.
\end{equation}
With the binary degree of freedom per site, the dimension of the Hilbert space (and thus the number of coefficients $c_{\sigma_1 \ldots \sigma_L}$) is $2^L$, so that the full diagonalization of the problem is exponentially expensive in terms of computational effort. Certain gains can be made by restricting the Hilbert space to, say, a specific spin or particle number sector \cite{Pietracaprina2018a}, but the exponential scaling is unavoidable and eventually makes it prohibitively expensive to compute the eigenspectrum beyond $L \ \approx 22$ and dynamics beyond $L \approx 28$. MPS-based algorithms avoid this problem by restricting the number of the coefficients, at the cost of approximating the exact many-body state, in the following way (for open boundary conditions):
\begin{equation}
|\Psi \rangle = \sum_{\{\sigma_i\}} A^{\sigma_1} A^{\sigma_2} \cdots A^{\sigma_L} \, |\sigma_1, \ldots, \sigma_L \rangle. \label{eq:mpsansatz}
\end{equation}
Here, each $A^{\sigma_i}$ is a matrix (whose dimension can in principle depend on $i$), whence the name \emph{matrix product state} emerges. In the case of open boundary conditions, the left and right matrices in the product are row and column vectors respectively, so that a scalar coefficient is obtained, while in the case of periodic boundary conditions one typically takes the trace of the product of matrices of fixed dimensions. The name \emph{matrix product states} is a bit of a misnomer:  we are in fact talking about a tensor as there are three indices per $A$. Mathematicians therefore prefer the name \emph{tensor train}. In principle, the full many-body state can, without loss of generality, be decomposed in such an MPS. We seem to have gained nothing, but the main benefit of this approach lies in a key approximation that can be made with the help of singular value decompositions. The result of this approximation is that the dimension of each matrix $A^{\sigma_i}$ is restricted to be (at most) $\chi \times \chi$, where $\chi$ is the bond dimension. During the singular value decomposition, we can discard the smallest singular values, retaining only the $\chi$ largest ones. As it turns out, this approximation corresponds physically to considering only a low-entanglement subsector of the Hilbert space.

What makes the MPS approach to be especially suitable for studying the physics around the MBL transition? As we have pointed out in Sec.~\ref{sec:avalanche}, it is expected on the basis of analytical arguments that the critical point is essentially localized and therefore exhibiting low entanglement. By continuity, this is expected to be applicable also to moderately large systems on the ergodic side close to the MBL transition. Such low-entanglement states are well-captured by tensor networks like MPS.

\subsection{Algorithms}

The MPS Ansatz \eqref{eq:mpsansatz} describes the wave function, but it does not prescribe how the coefficients ought to be obtained. For this purpose there are two classes of algorithms: those targeting statics (the eigenstates of the system and their properties) and those targeting dynamics starting from some arbitrary state.

Lim and Sheng \cite{Lim2016a} considered the MBL problem using DMRG-X, an algorithm based on White's original DMRG algorithm \cite{White1992a} which targets ground states. DMRG-X instead targets highly excited states. Normally, DMRG would be unable to distinguish such states since the level spacing becomes too small. DMRG-X exploits the fact that, on the localized side of the transition, different eigenstates differ in both energy and spatial extent, boosting the resolution of the algorithm. In this way, the authors of Ref.~\cite{Lim2016a} were able to capture the MBL transition numerically. An alternative quantum-circuit approach that can be applied to both 1D and 2D systems to target highly excited eigenstates was pioneered by Wahl \emph{et al.}~\cite{Wahl2017a, Wahl2019a}, building on an earlier work by Bauer and Nayak \cite{Bauer2014a}.

In this review, we mainly focus on MPS investigation of quantum dynamics. Early studies of MBL using MPS \cite{Znidaric2008a, Bardarson2012a} employed time-dependent DMRG (tDMRG) or a closely related method of time-evolving block decimation (TEBD) \cite{Vidal2003a, Daley2004a, White2004a}. The limiting factor in terms of the applicability of the method is the accumulation of errors at each time step. As one crosses the MBL transition from the MBL side to the ergodic side, the entanglement grows with time faster. Therefore, for a given bond dimension $\chi$ and desired error $\epsilon$, the maximum time that can be reached decreases with decreasing disorder strength $W$. For TEBD and t-DMRG, at each time step there is a \emph{truncation error} that results from repeated truncations of the entanglement spectrum.

Recently, a novel MPS-based algorithm was proposed, which does not suffer from these truncation errors: the \emph{time-dependent variational principle} (TDVP) \cite{Haegeman2016a}. Instead of a truncation, there is a \emph{projection} onto the variational subspace described by the MPS:
\begin{equation}
\frac{\partial}{\partial t} |\psi \rangle  = -i \mathcal{P}_\mathrm{MPS}\mathcal{H}|\psi\rangle, \label{eq:tdvp}
\end{equation}
where $\mathcal{H}$ is the Hamiltonian of the system and $\mathcal{P}_\mathrm{MPS}$ is the projection operator. Of course, it is not a priori given that the projection errors induced by the time evolution according to Eq.~\eqref{eq:tdvp} are smaller than those induced in other MPS methods. In fact, this depends on the problem at hand \cite{Paeckel2019a}. For the MBL problem, the TDVP approach turns out to be advantageous as will be discussed in more detail below. This is related to the fact that the projection conserves the globally conserved quantities, in particular, the energy. The MPS-TDVP approach permits to controllably explore the time evolution up to sufficiently long times, $t \sim 100$ -- 300 in units of the lattice hopping times for disorder $W$ around $W_c$. This is the same time range as the one accessed in cold-atom experiments.

Tensor networks permitting to tackle problems beyond 1D geometry have also been developed. In particular, an approach based on projected entangled pair-states (PEPS) has been applied to the 2D  MBL problem \cite{Kennes2018a, Kshetrimayum2019a}. It is also possible to generalize the MPS ansatz to different lattice geometries by introducing longer-range terms in the Hamiltonian, by using matrix product operators (MPOs) \cite{Verstraete2004a, Pirvu2010a}. It is not a priori clear which approach is the most suitable for the MBL problem beyond 1D geometry. As we discuss below, however, the MPS-TDVP-based approach provides a particularly efficient method of simulation also in this situation.

A distinct MPS-based approach for simulating dynamics was recently developed by \v{Z}nidari\v{c} \emph{et al.}~\cite{Znidaric2016a}. Their approach uses the Lindblad master equation to simulate the current driven through the system. In a certain sense, this approach is the ``opposite'' of the usual MPS-based approaches as it works most efficiently in the ergodic limit of \emph{low} disorder. The authors of Ref.~\cite{Znidaric2016a} have successfully used this approach to capture the crossover from diffusive behaviour at very weak disorder to subdiffusive behaviour at moderate disorder. However, as the critical point has essentially the localized character, this method loses its efficiency when one moves towards $W_c$ from the ergodic side. Specifically, as one approaches the MBL transition, the time scales required for the current to saturate diverge.

\section{Model and observables}

\subsection{Model}

We will consider the following models: (i) a 1D chain with random disorder, (ii) a 1D quasiperiodic chain, and (iii) a 2D square lattice with random disorder.

\subsubsection{One dimension, random disorder} 
\label{sec:1drandommodel}
\label{sec:model-1D}

We first discuss the most widely studied model: the XXZ Heisenberg chain of length $L$ with on-site random disorder:
\begin{equation}
\mathcal{H} = \sum_{i} \left[ \frac{J}{2}\Big(S_i^+S_{i+1}^- + S_i^- S_{i+1}^+ \Big) + \Delta S_i^z S_{i+1}^{z} + h_i S_i^z\right],
\label{eq:xxzham}
\end{equation}
where $S_i$ are spin-$\frac{1}{2}$ operators, $\Delta/J$ is the anisotropy, and the on-site field $h_i$ is taken from a uniform distribution $h_i \in [-W, W]$, with $W$ denoting the strength of disorder. By virtue of Jordan-Wigner transformation, this Hamiltonian is equivalent to that of interacting fermions (or hard-core bosons):
%(neglecting terms at the boundary, which are irrelevant in the limit of large systems)
\begin{equation}
\mathcal{H} = \sum_{i} \left[ \frac{J}{2} (b_{i}^\dagger b_{i+1} + \mathrm{H.c.})  + \Delta \hat{n}_{i} \hat{n}_{i+1} + h_i \hat{n}_i \right].
\end{equation}
Here $b_i$, $b_{i}^\dagger$ are the annihilation and creation operators for a spinless fermion or a hard-core boson, $\hat{n}_i=b_i^\dagger b_i-1/2$, and $J$ and $\Delta$ are the hopping amplitude and the nearest-neighbor interaction, respectively. Typically, numerical studies choose $J=\Delta = 1$, although sometimes values of $\Delta/J$ different from unity are chosen in order to avoid possible peculiarities of the isotropic $\Delta/J = 1$ Heisenberg chain. However, with sufficiently strong disorder, those peculiarities should not be of importance. We will focus on the $J=\Delta = 1$ model below.

\subsubsection{One dimension, quasiperiodic potential}
\label{sec:model-1D-quasi}

An alternative, which has been investigated in experiments \cite{Schreiber2015a}, is to take a \emph{quasi-periodic} potential, which is not truly random. Specifically, one typically considers the many-body generalization of the Aubry-Andr\'e-Harper problem \cite{Harper1955a, Aubry1980a}, wherein the potential experienced by particles is given by
\begin{equation}
h_{i} = \frac{W}{2} \cos(2 \pi \Phi i + \phi_0), \quad \phi_0 \in [0, 2 \pi). \label{eq:AA_field}
\end{equation}
Here $W$ again represents the strength of the potential, while $\Phi$ is the period of the potential. If one chooses $\Phi$ irrational, then the potential periodicity is incommensurate with that of the lattice. Further, $\phi_0$ is a random phase; one averages over a uniform distribution of $\phi_0$ over $[0, 2\pi)$.

It is well known that the non-interacting ($\Delta=0$) Aubry-Andr\'e model undergoes a localization transition at $W=2$ (for all energies). Thus, for $W>2$ all states are localized in the absence of interaction, and one can expect a MBL transition when the interaction is switched on, in analogy with the random-potential setting. However, despite clear similarities, there are also qualitative differences between the random and quasiperiodic problems. In particular, while a random potential has extended rare regions where the potential is anomalously weak or anomalously strong, a quasiperiodic potential cannot have such regions. Therefore, one should expect qualitative differences in the behaviour of those observables that are essentially related to rare events. On the experimental side, quasiperiodic systems are easier to study, at least in cold-atom experiments, which is why the earliest MBL experiments \cite{Schreiber2015a} considered this type of systems.

\subsubsection{Two dimensions, random disorder}
\label{sec:model-2D}

The Hamiltonian, in the hard-core boson language of section \ref{sec:1drandommodel}, can be straightforwardly generalized to a 2D (square) lattice:

\begin{equation}
\mathcal{H} =  \sum_{\langle ij;i'j' \rangle} \left[ -\frac{J}{2} (b_{ij}^\dagger b_{i'j'} + \mathrm{H.c.})  + U\hat{n}_{ij} \hat{n}_{i'j'} \right] + \sum_{ij} h_{ij} \hat{n}_{ij}.
\label{eq:ham}
\end{equation}
Here $b_{ij}^\dagger$ is again the creation operator for a hard-core boson at the site given by indices $i,j$, $U$ is the nearest-neighbor interaction, $J$ is the energy associated with particle hopping to neighboring sites, and $h_{ij}$ is the magnitude of the on-site potential, with $\hat{n}_{ij} \equiv b_{ij}^\dagger b_{ij}$. We define this model on a square lattice of size $L \times d$, with $i \in [1, L]$ and $j \in [1, d]$.

\subsection{Imbalance}

Numerical studies---as well as experiments---often consider the density \emph{imbalance}. It is used as the key observable characterizing the dynamics in the MPS-TDVP data presented below.

For a 1D geometry, the idea is that one initially prepares the system in a charge-density-wave (N\'eel) state:
\begin{equation}
\psi(t = 0) = \{ 1, 0, 1,0, \ldots, 1, 0 \},
\end{equation}
where 0 or 1 indicates the local occupation and corresponds to, respectively, $-1/2$ or $1/2$ in the $S^z$-basis in the spin language. The imbalance is then defined as the memory of this state at later times:
\begin{equation}
\mathcal{I}(t) =  \frac{n_\mathrm{odd} - n_\mathrm{even}}{ n_\mathrm{odd} + n_\mathrm{even}} \,,
\end{equation}
where $n_\mathrm{odd}$  and $n_\mathrm{even}$ are the densities at odd and even sites, respectively. By definition,  $\mathcal{I}(0) = 1$.
We will also use a generalization to the 2D case, wherein initially \emph{columns} (fixed $i$) will  be taken to be occupied for odd $i$ and unoccupied for even $i$.

The appeal of the imbalance lies in the fact that, first, it is relatively easily accessed in experiments and, second, it relaxes to zero rapidly in the absence of disorder. Further, the density can be easily computed using MPS approaches \cite{Schollwock2011a}, as opposed to the level statistics of eigenstates, which requires exact diagonalization \cite{Luitz2015a}.

Inspection of the decay of the imbalance provides a way for determining the transition from the ergodic to the localized phase.
In the ergodic phase, $W < W_c$, one expects a power-law decay,
\cite{Luitz2017a}:
\begin{equation}
\mathcal{I}(t) \propto t^{-\beta} \,.
\label{I-t-beta}
\end{equation}
On the other hand, in the MBL phase, $W > W_c$  (and at  the critical disorder $W = W_c$), one expects $\beta = 0$ corresponding to the long-time saturation of the imbalance at a non-zero value. Of course, the imbalance  $ \mathcal{I}(t)$, as obtained in finite-time simulations, does not exactly follow the power-law decay \eqref{I-t-beta}. One can generalize Eq.~\eqref{I-t-beta} by defining a running power-law exponent $\beta(t) = - \partial \ln  \mathcal{I}(t) /\partial \ln t$. One then expects $\lim_{t \rightarrow \infty} \beta(t) = 0$ for $W > W_c$  and $\lim_{t \rightarrow \infty} \beta(t) > 0$ for $W < W_c$. (Of course, the $t \to \infty$ limit should be taken after the limit of an infinite system size.) Clearly, when the position of $W_c$ is extracted from finite-time simulations, one has to rely on an extrapolation to long times. Let us note that not only the limiting value of $\beta$ at $t \to \infty$ but also the whole time dependence $\beta(t)$ is of interest, since it contains important information about the dynamics of many-body localization and delocalization. It is also worth emphasizing that, since the characteristic times in MPS-TDVP numerical simulations presented below are of the same order as in experiments, one can directly compare $\beta(t)$ to experimental measurements.

\subsection{Entanglement entropy}

Another popular measure for the analysis of MBL-type systems is the \emph{entropy} $S$, by which one usually means the bipartite von Neumann entropy of entanglement \cite{Laflorencie2016a}. The entropy corresponding to a subsystem $A$ of the system can be obtained by tracing out the degrees of freedom of the exterior $B$:
\begin{equation}
S_A = -\mathrm{Tr}(\rho_A \ln \rho_A), \quad \rho_A \equiv \mathrm{Tr}_B |\Psi \rangle \langle \Psi |.
\end{equation}
Most commonly, the bipartite entropy for a division of the system in two halves is used. This quantity, too, can be obtained straightforwardly by MPS methods---in fact, it is obtained ``for free'' during the course of MPS algorithms \cite{Schollwock2011a}. The entropy grows logarithmically in the localized phase \cite{Bardarson2012a}, and faster than logarithmically on the ergodic side. Because a logarithmic growth is difficult to distinguish from a weak power-law growth, this makes the transition more difficult to pinpoint using entropy-related measures than using measures derived from particle transport---which freezes on the MBL side of the transition. For this reason, we will focus on particle transport---quantified by the imbalance---in the remainder of this review.

\section{Numerical results from matrix product states}

We will now present the results for the MBL problem as obtained from MPS simulations.

\subsection{One dimension, purely random disorder}

We begin with the case of a random 1D system. The model defined on lattice of length $L$ (with sites labeled by $i = 1, \ldots, L$) was defined in Sec.~\ref{sec:model-1D}. This model (or equivalent models) was recently considered using the TDVP \cite{Doggen2018a, Kloss2018a, Chanda2019a}. In Ref.~\cite{Kloss2018a}, Kloss \emph{et al.}~argue that the TDVP can be used to target times up to $t \sim 100$ even for moderate disorder $W = 1.5$. This value is significantly below the estimate for the critical disorder $W_c \approx 3.7$ \cite{Luitz2015a} as obtained from exact diagonalization, suggesting that the TDVP is efficient also rather deep into the ergodic regime.

%%%%%%%%%%%%%%%%%
\begin{figure}
    \centering
    \includegraphics[width=.8\columnwidth]{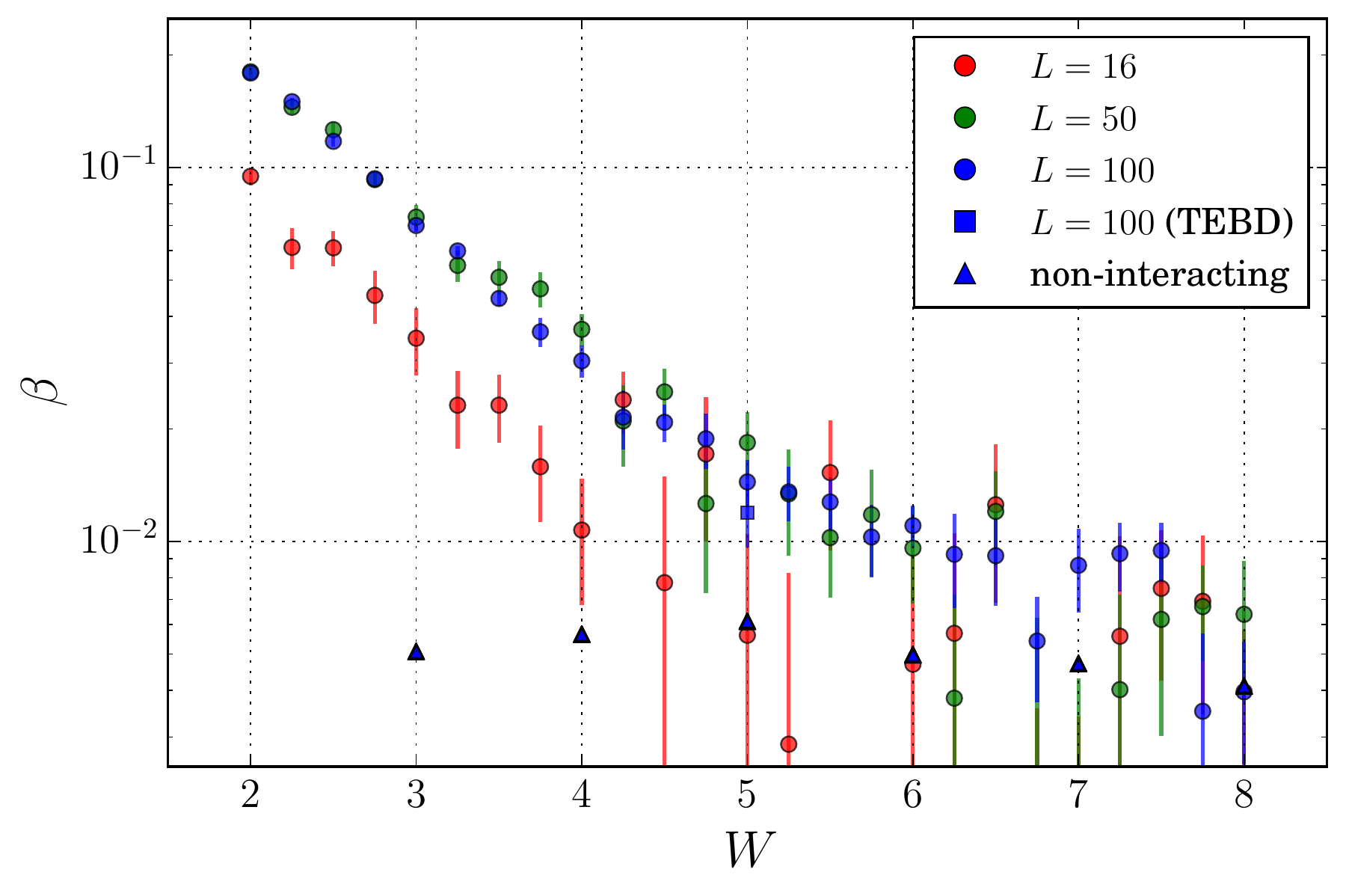}
    \caption{Decay of the imbalance exponent $\beta$ as a function of the disorder strength $W$ for system sizes $L=16$, 50, and 100, as computed from the time window $t \in [50, 100]$.
    For comparison, data for a non-interacting system are also shown.
    From Ref.~\cite{Doggen2018a}, \textcopyright ~2018 American Physical Society.}
    \label{fig:MBLbeta1D}
\end{figure}
%%%%%%%%%%%%%%%%%%

In Ref.~\cite{Doggen2018a}, by the present authors together with Schindler, Tikhonov, and Neupert, the slow transport around the MBL transition was explored and the position of the transition was estimated by the MPS-TDVP method, in systems up to $L=100$ and at times up to $t=100$.  Numerical results from Ref.~\cite{Doggen2018a} are shown in Fig.~\ref{fig:MBLbeta1D}. We can observe that $\beta(W)$ is a monotonically decreasing function of $W$, as expected. In the case that the system is localized, $\beta(W > W_c) \rightarrow 0$ as $t \rightarrow \infty$. As was already mentioned above, since only finite times are available, an extrapolation is needed, and the error made during the course of this extrapolation has to be estimated. In Ref.~\cite{Doggen2018a}, this error is taken as bounded from below by the residual decay of the non-interacting system (triangle symbols in Fig.~\ref{fig:MBLbeta1D}). The critical disorder $W_c$ is then estimated by adding statistical errors resulting from the finite number of different disorder realizations. This leads to the estimate $W_c \approx 5.5$.

It is important to emphasize that the data for $L = 16$ yield, within the same procedure, a substantially smaller estimate, $W_c \approx 4$, which is in good agreement with the results of exact diagonalization for systems of comparable size ($W_c \approx 3.7$ in Ref.~\cite{Luitz2015a}  and $W_c \approx 4.2$ in Ref.~\cite{PhysRevResearch.2.042033}). This agreement with the exact diagonalization for small systems serves as an important benchmark for the estimation of $W_c$ on the basis of imbalance dynamics. As the system size increases from $L=16$ to $L=50$, a substantial drift of $\beta$ is observed, leading to a substantially higher value of $W_c \approx 5.5$. This drift towards delocalization with increasing system size is similar to that found in the RRG model but is still more pronounced. Indeed, in the present case of a spin chain, one expects an additional contribution to the drift that originates from rare ergodic spots discussed in Sec. \ref{sec:avalanche}.
At the same time, the values for $L = 50$ and $L = 100$ agree within error bars, indicating that system sizes of $L \approx 50$ are sufficient to essentially reach the $L \to \infty$ saturation. This suggests that the estimate $W_c \approx 5.5$ for the critical disorder obtained in Ref.~\cite{Doggen2018a} does correspond to the thermodynamic limit $L \to \infty$. This estimate was further supported in Ref.~\cite{Doggen2018a} by a machine learning analysis \cite{Carleo2019a} of the data obtained from the TDVP, which yielded essentially the same value of $W_c$.

Chanda \emph{et al.} studied the same problem using both TDVP and time-dependent DMRG \cite{Chanda2019a}. They considered longer times (up to $t = 500$ for $L=50$) and characterized the decay of the imbalance by $\beta(t)$. The data of Ref.~\cite{Chanda2019a} support the conclusion of Ref.~\cite{Doggen2018a} about a substantial drift of the critical point towards stronger disorder with increasing $L$. The authors of Ref.~\cite{Chanda2019a} used, however, a significantly higher value for the error, $\beta_\mathrm{error} = 0.02$, than the error estimate of Ref.~\cite{Doggen2018a}, thus attributing all values $\beta <  \beta_\mathrm{error}$ to the localized phase. Clearly, increasing the cutoff $\beta_\mathrm{error}$ reduces the estimate for $W_c$. As a result, Ref.~\cite{Chanda2019a} came to an estimate $W_c \approx 4.2$.  It is worth pointing out that a very recent work of the same group, using an advanced version of the exact-diagonalization approach for up to $L=24$ in combination with an extrapolation to $L \to \infty$, obtained the estimate $W_c \approx 5.4$ \cite{Sierant2020b}, which is in perfect agreement with the result of Ref.~\cite{Doggen2018a}.

Prior to the above TDVP studies, Hauschild \emph{et al.}~\cite{Hauschild2016a} considered the same model by tDMRG.  Inspired by the experiment \cite{Choi2016a}, they used a different initial state characterized by  a domain wall between two oppositely polarized halves of the system (empty and filled, in the fermionic language), and studied the dynamics of the domain wall melting.   Qualitatively, the results are in agreement with those found for the N\'eel state. At the same time, the authors of Ref.~\cite{Hauschild2016a} did not attempt an accurate analysis of transport close to the transition point, of the value of $W_c$, and of its drift with the system size.

%%%%%%%%

Recent works \cite{Suntajs2020a, Sels2020a} have emphasized a sizeable drift of the critical point $W_c(L)$ as found in exact diagonalization (i.e. for $L \lesssim 20$) of the 1D model  \eqref{eq:xxzham}.
Specifically, their results for the drift can be cast in the form  $\partial W_c / \partial L \approx 0.1$ at these length scales. On this basis, the authors of Refs.~\cite{Suntajs2020a, Sels2020a} have cast doubt on whether $W_c$ stays finite in the $L \to \infty$ limit in this model. It is worth mentioning that a hypothesis of $W_c$ growing linearly with $L$ is in dramatic disagreement with the existing analytical results on the MBL transition. Indeed, such a fast growth cannot be related to any kind of rare events; if for real, it should be seen on the level of the perturbative treatment. However, no such analytical evidence exists, see Sec.~\ref{sec:pert}. We believe that the numerical results  of Refs.~\cite{Suntajs2020a, Sels2020a} actually do not provide a ground for the conclusion on the asymptotic growth of $W_c$; they rather demonstrate strong finite-size corrections. Indeed, the  magnitude of the drift is in very good agreement with our result that $W_c$ reaches the value of $W_c \approx 5.5$ for $L=50$. Of course, using only the exact-diagonalization numerical data (which correspond necessarily to a relatively narrow range of quite small $L$), it is extremely difficult to judge on a saturation of this drift at large $L$. It was emphasized in Refs.~\cite{Abanin2019a, Sierant2020a} that the results of \v{S}untajs \emph{et al.}~are due to finite-size effects and that a similar, approximately linear, drift of $W_c$ is observed at small $L$ also in models for which it is known analytically that $W_c$ is finite in the imit $L \to \infty$. An important example is the RRG model discussed above. It was shown in Ref.~\cite{Sierant2020b} that an assumption of $1/L$ finite-size effects yields a good fit to the exact-diagonalization data, resulting in the estimate $W_c \approx 5.4$. Moreover, a recent work by Panda \emph{et al.}~\cite{Panda2020a} argues that system sizes of at least around $L \approx 50$ are needed to assess the existence of MBL. This perfectly agrees with our finding that a substantial drift of the apparent $W_c$ is observed between $L \approx 20$ and $L =50$, while the system sizes  $L = 50$ and $L = 100$ yield nearly identical results.

\subsection{One dimension, quasiperiodic systems}
\label{sec:quasi}

%%%%%%%%%%%%%%%%%
\begin{figure}
    \centering
    \includegraphics[width=0.8\columnwidth]{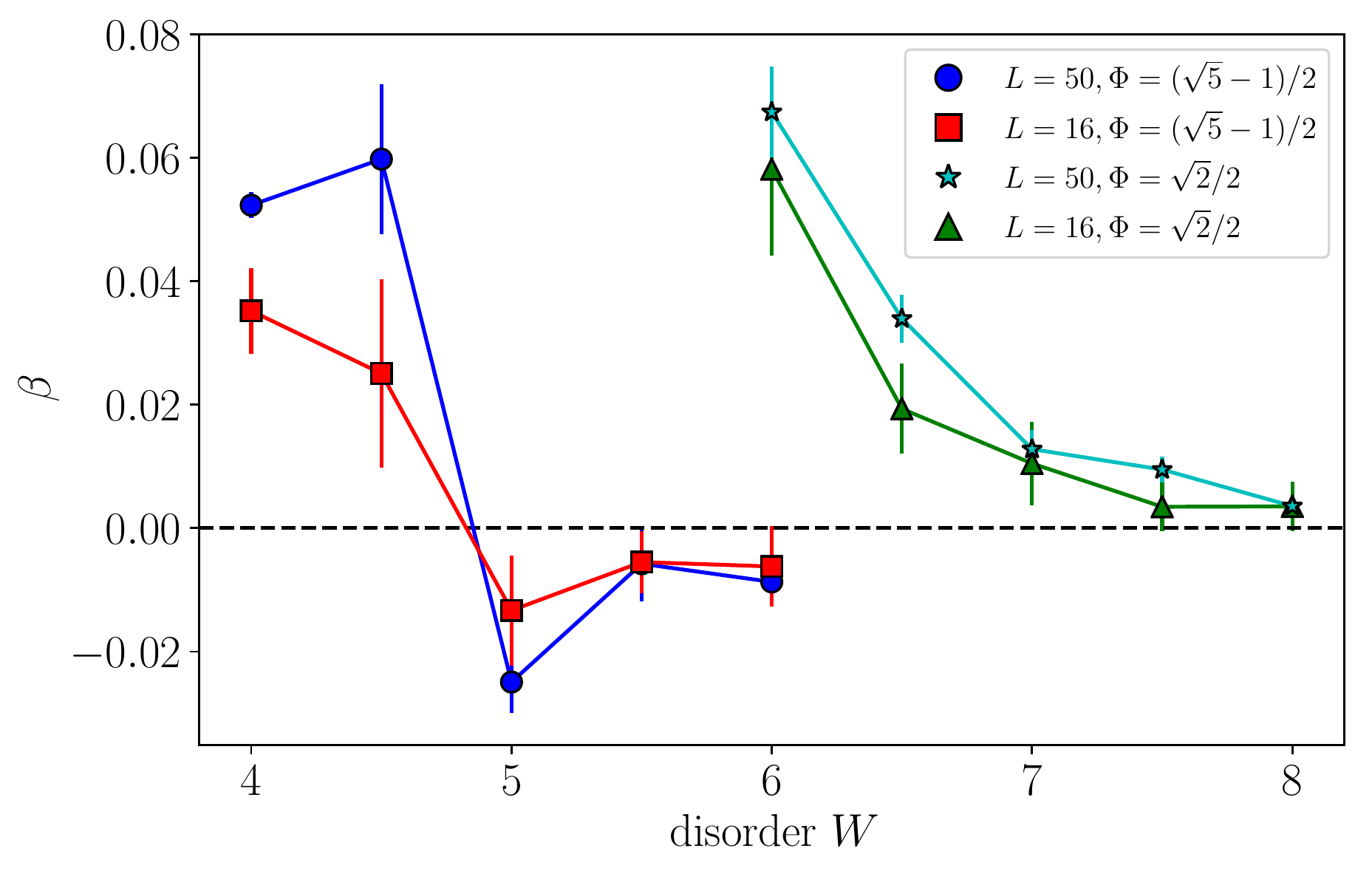}
    \caption{Imbalance exponent $\beta$ in the case of the quasiperiodic potential \eqref{eq:AA_field} for various system sizes $L$, field strengths $W$ and periodicity $\Phi$, over the time window $t \in [50, 180]$.  Results are averaged over $\approx 400)$ random choices of the phase $\phi_0$. Figure adapted from Ref.~\cite{Doggen2019a}, \textcopyright ~2019 American Physical Society.}
    \label{fig:AAbeta1D}
\end{figure}
%%%%%%%%%%%%%%%%%%

We now consider the case of a quasiperiodic potential \eqref{eq:AA_field}. This problem was studied, using the TDVP-MPS approach, in Ref.~\cite{Doggen2019a}, for system sizes up to  $L=50$ and times up to $t=300$. The results are shown in Fig.~\ref{fig:AAbeta1D}. As in the truly random case discussed above, the results for the imbalance dynamics provide a clear evidence of the MBL transition and allow one to estimate $W_c$ in a large system.

One important result of Ref.~\cite{Doggen2019a} is a strong dependence of the critical disorder $W_c$ on the periodicity $\Phi$. In particular, for two values of $\Phi$, data for which are shown in Fig.~\ref{fig:AAbeta1D}, the results are $W_c \approx 4.8$ for $\Phi = (\sqrt{5}-1)/2$ and $W_c \approx 7.8$ for $\Phi = \sqrt{2}/2$. This strong dependence can be qualitatively explained as resulting from a difference in the statistics of the potential at neighbouring sites \cite{Doggen2019a, Guarrera2007a}.  This affects properties of localized single-particle states, which in turn influence susceptibility of the system to interaction-induced many-body delocalization.

While there are similarities in the qualitative behaviour of $\beta(W)$ in the random and quasiperiodic cases, there are also significant differences.
First, the finite-size effects are substantially smaller in the quasiperiodic case, as compared to the purely random case: the estimate of $W_c$ depends only weakly on the system size. This is in agreement with the theoretical predictions \cite{Khemani2017a}, where it is argued that the disordered and quasiperiodic cases are essentially different. Second, the data on the ergodic side of the MBL transition in a quasiperiodic system indicate an increase of $\beta(t)$ with time (see Fig.~\ref{fig:mps_compare}). On the time scale probed by MPS-TDVP this increase is rather modest (see Sec.~\ref{sec:comparison-MPS}), in agreement with experiments that reported a slow decay that can be well fitted by a power law. At the same time, the increase of $\beta(t)$ is consistent with the absence of the effects of rare regions for the quasiperiodic potential, in contrast to the case of a random potential, where these effects are expected to lead to subdiffusive transport in the long-time limit, as was discussed above.
 Recent studies by time-dependent Hartree-Fock approximation \cite{Weidinger2018a} (that allows one to reach much longer times at the expense of losing the full control of accuracy provided by the MPS-TDVP method) confirm the qualitative difference in the behaviour of $\beta(t)$ for random and quasi-periodic systems.

\subsubsection{Comparison of MPS methods}
\label{sec:comparison-MPS}

%%%%%%%%%%%%%%%%
\begin{figure}
    \centering
    \includegraphics[width=0.7\columnwidth]{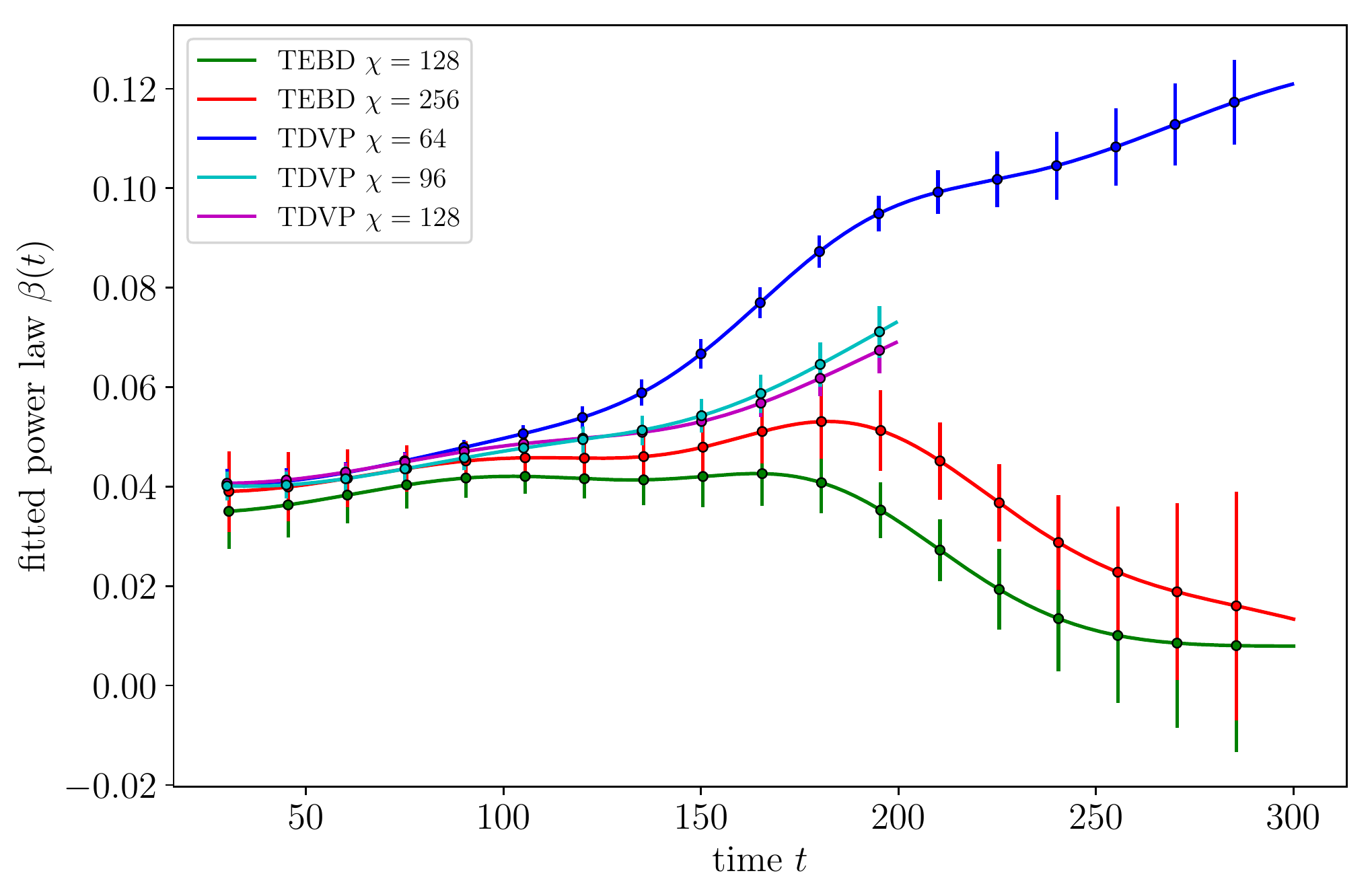}
    \caption{Comparison  of the imbalance decay exponent $\beta(t)$ using two different MPS methods, TDVP and TEBD, using various choices of the convergence parameter $\chi$. Parameters are: the strength of the quasiperiodic field $W = 4$, system size $L=50$, and periodicity $\Phi = (\sqrt{5}-1)/2$. Different disorder realizations were used for each choice. From Ref.~\cite{Doggen2019a}, \textcopyright ~2019 American Physical Society.}
    \label{fig:mps_compare}
\end{figure}
%%%%%%%%%%%%%%

We have also compared \cite{Doggen2019a}, in the quasiperiodic case, the time-dependent behaviour of the imbalance decay, quantified using $\beta(t)$, for two different MPS methods: the TDVP and TEBD. The results are shown in Fig.~\ref{fig:mps_compare}. The following qualitative features are observed. First of all, and most reassuringly, we observe a clear convergence of both methods to the same curve with increasing bond dimension $\chi$. Second, the convergence with $\chi$ is superior in the case of the TDVP (at least from the perspective of the imbalance dynamics): at fixed time $t \approx 200$, the difference between $\chi = 96$ and $\chi = 128$ in TDVP is smaller than the deviation of the TEBD data computed with even higher bond dimension $\chi = 256$. There is also another qualitative difference between the two methods: in the case of TDVP, when convergence is lost, the algorithm appears to overestimate the decay, whereas in the case of TEBD the opposite is true. Specifically, the TEBD results show a downturn in $\beta(t)$ towards zero reflecting the spurious saturation of the imbalance. With increasing $\chi$, the time at which this downturn starts becomes longer. The superior performance of TDVP with respect to TEBD on the ergodic side of the MBL transition was attributed  \cite{Doggen2019a} to different ways of truncation and, in particular, to non-conservation of energy within TEBD.  A related analysis was performed in Ref.~\cite{Chanda2019a} for the truly random case, with similar findings.

In the case of TDVP, convergence is reached up to $t \approx 180$ using $\chi = 128$. Up to that time scale, a modest increase in $\beta$ of roughly $20 \%$ is observed, whereas in the case of random disorder, the power law appears to be robust over such time scales, at least on the ergodic side of the transition. This difference between the quasiperiodic and random cases was already pointed out in Sec.~\ref{sec:quasi}.

\subsection{From one dimension to two dimensions: Quasi-1D and 2D random systems}

\begin{figure}
    \centering
    \includegraphics[width=0.8\columnwidth]{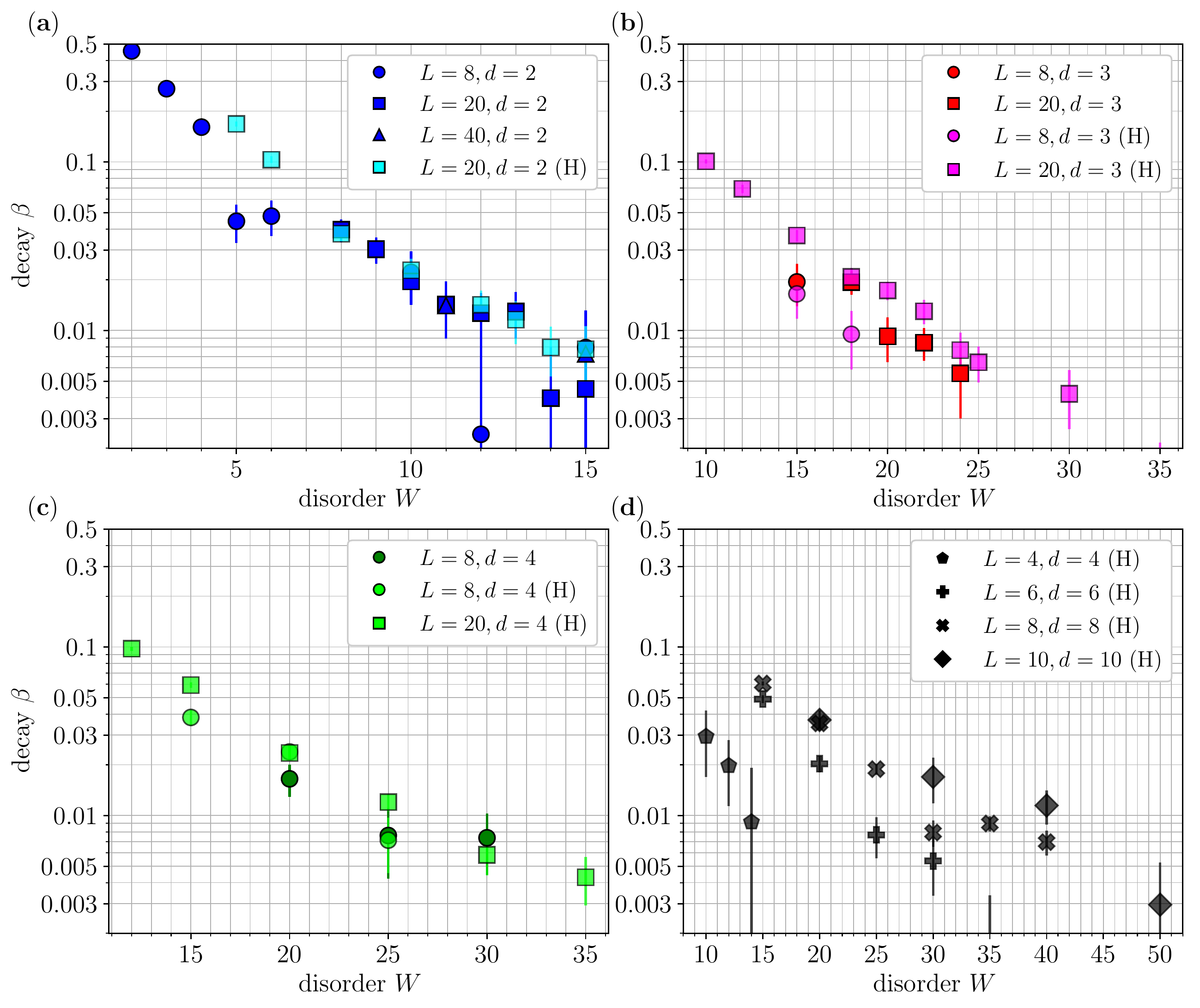}
    \caption{Imbalance decay exponent $\beta$ for quasi-1D and 2D systems with random disorder, with length $L $ and width $d$, and various disorder strengths $W$, as computed for the time window $t \in [50,100]$. $\mathrm{a)}$ two-leg ladder, $d=2$, $\mathrm{b)}$ three-leg ladder, $d=3$, $\mathrm{c)}$ four-leg ladder, $d=4$, $\mathrm{d)}$ 2D case, $d=L$. Different colours in the same panel indicate different implementations of the TDVP. From Ref.~\cite{Doggen2020a}, \textcopyright ~2020 American Physical Society.}
    \label{fig:beta2D}
\end{figure}

\begin{figure}
    \centering
    \includegraphics[width=0.8\columnwidth]{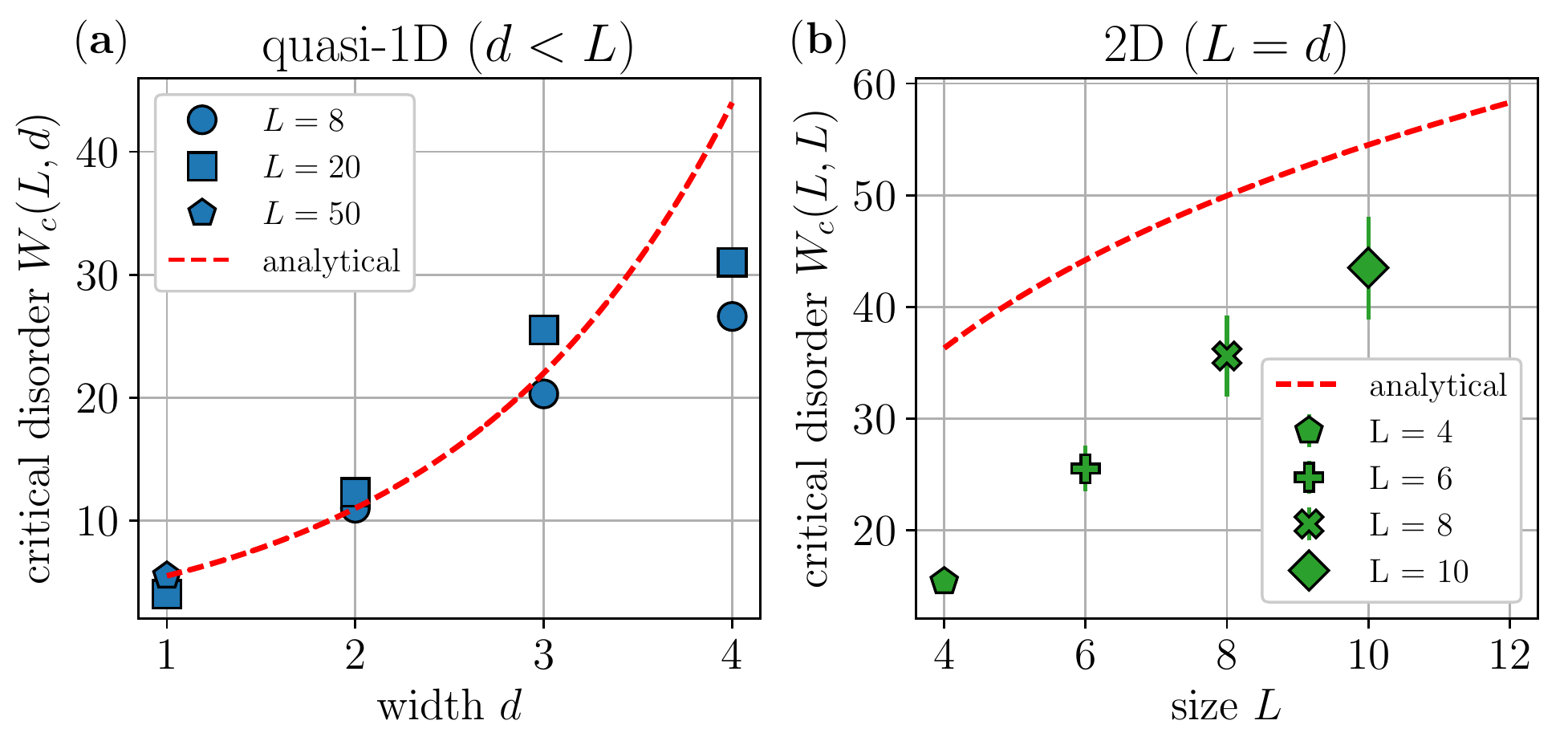}
    \caption{Critical disorder $W_c(L,d)$ for $(\mathrm{a)}$ quasi-1D and $(\mathrm{b)}$ 2D cases, compared to analytical predictions for
    (a) $W_c(\infty,d)$ with $d \gg 1$ and (b) $W_c(L,L)$ with $L \gg 1$ (see Sec.\ref{sec:avalanche}). From Ref.~\cite{Doggen2020a}, \textcopyright ~2020 American Physical Society.}
    \label{fig:Wc2D}
\end{figure}

In Ref.~\cite{Doggen2020a}, the MPS-TDVP approach was extended to study the model on a 2D square lattice, see Sec.~\ref{sec:model-2D}. More specifically, two types of geometry were investigated: quasi-1D ladders of $L \times d$ sites, with $d=2, 3,$ and 4, and $L \gg d$, and 2D samples with $d=L$. The model is determined by the Hamiltonian \eqref{eq:ham}, with $i \in [1,L]$ and $j \in [1,d]$. In the initial state, \emph{columns} with odd index $i$ are occupied, similar to the experiment \cite{Rubio-Abadal2019a}. The results for the exponent $\beta$ of the columnar imbalance decay are shown in Fig.~  \ref{fig:beta2D}. In analogy with the 1D case, they were used to determine the critical disorder $W_c(L,d)$. The resulting estimates are shown in Fig.~\ref{fig:Wc2D} for the quasi-1D (left panel) and 2D (right panel) cases, in comparison with analytical predictions based on the avalanche theory, Eq.~\eqref{eq:q1dWc}.

%%%%%%%%%%%%
%%%%%%%%%%%%%%

%%%%%%%%%%%%%%
%%%%%%%%%%%%%%

In the quasi-1D case, the second line of Eq.~\eqref{eq:q1dWc} predicts an exponential increase of $W_c(L,d)$ with increasing number of legs $d$ in the ladder, $W_c(L,d) \sim 2^d$, under the condition $L > L_*(d)$, where $L_*(d)$ is given by Eq.~\eqref{eq:L-star}. Importantly, $L_*(d)$ increases very fast with $d$ according to this formula: it is rather small for $d=1,2,$ and 3, is moderately large for $d=4$ [$ L_*(4) \approx 60$], and is very large for $d \ge 5$. The numerical results in the left panel of Fig.~\ref{fig:Wc2D} are in very good agreement with these predictions: at $d=4$, the increase of $W_c(L,d)$ deviates downwards from the exponential curve [corresponding to the large-$L$ limit of Eq.~(\ref{eq:q1dWc})].

In the 2D case, the analytically expected increase of $W_c(L,L)$ is given by the first line of Eq.~\eqref{eq:q1dWc}. While it is asymptotically slower than a power law, it yields a rather big increase over the available interval of system sizes from $L=4$ to $L=8$. As is seen in the right panel of Fig.~\ref{fig:Wc2D} the values of $W_c(L,L)$ provided by the TDVP analysis do show a strong increase, which is in fact even somewhat faster. The agreement is rather good, taking into account that the system sizes $L$ are moderately large, while the formula has been derived for the asymptotic limit of large $L$.

Hauschild \emph{et al.}~\cite{Hauschild2016a} have earlier considered this model in the case of a two-leg ladder ($d=2$) by using the tDMRG approach and the domain-wall melting as a probe of MBL. Their data are in general agreement with the above results. While they estimate the critical disorder for $d=2$ and large $L$ as $8 \lesssim W_c \lesssim 10$, their Fig. 4 shows that in fact the dynamics of melting has not saturated, either for $W=8$ or for $W=10$. Thus, the data of Ref.~\cite{Hauschild2016a} indicate a critical disorder $W(\infty, 2)$ higher than 10,  in consistency with the result $W_c(\infty,2) \approx 13$ of Ref.~\cite{Doggen2020a}.

\section{Summary}

In this article, we have reviewed recent results on application of the MPS-based approach to the investigation of quantum dynamics in interacting disordered systems around the MBL transition. The focus was put on the advances achieved in the framework of the MPS-TDVP method \cite{Doggen2018a, Doggen2019a, Doggen2020a} and a comparison to the results obtained by other methods. The most salient conclusions of these studies are as follows:

\begin{enumerate}

\item The MPS-based framework, and in particular the MPS-TDVP approach, is a powerful tool for the investigation of MBL, complementary to exact diagonalization. It allows one to study controllably large systems, with $\sim$ 50 -- 100 ``qubits,'' up to times $t \sim 300$ close to the critical region. These system sizes and time scales are essentially the same as probed in the state-of-the-art experiments. It is demonstrated that the method works successfully for 1D, quasi-1D, and 2D  random systems as well as 1D quasi-periodic systems. It is shown that the imbalance decay, which is also studied experimentally and can be characterized by a power-law exponent $\beta$, is a convenient observable for monitoring the dynamics and locating the MBL transition within the MPS approach.

\item On the explored time scales, the systems show a slow, subdiffusive transport in a rather broad range of the disorder strength $W$ on the ergodic side of the MBL transition. This agrees with analytical expectations with respect to the localized character of the critical point of the MBL transition and of the effect of rare regions. In connection with this, one finds also a relatively slow growth of the entanglement in a sufficiently broad window around the transition point. This is highly favorable for the MPS-based approaches: one obtains convergence with modest values of the bond dimension $\chi$ up to rather long times.

\item For 1D random spin chains, which serve as a ``standard model'' of the MBL transition, the MPS study demonstrates a substantial drift of the critical point $W_c(L)$ with the system size $L$. This drift is similar to that found for the RRG model but is even more pronounced, which can be attributed to the effect of rare regions. Specifically, while for $L \approx 20$  we find $W_c \approx 4$, as also given by exact diagonalization, the MPS-TDVP results for $L = 50$ and $L=100$ yield $W_c \approx 5.5$.  The agreement with exact diagonalization for small system size corroborates the validity of the determination of the critical point on the basis of quantum dynamics at intermediate time scales. The fact that the values of $W_c$ obtained for $L=50$ and $L=100$ systems are nearly equal suggests that $W_c \approx 5.5$ can be considered as an estimate for the thermodynamic limit $L \to \infty$.

\item For quasi-periodic (Aubry-Andr\'e) systems, the finite-size effects are much weaker, which is consistent with the absence of rare regions with anomalously large or weak disorder in the quasi-periodic setting. Further, a growth of $\beta(t)$ with $t$ is found in the quasi-periodic case, again consistent with the difference in comparison with random system in what concerns the rare regions. The MBL transition point $W_c$ is found to depend strongly on the period $\Phi$ of the quasi-periodic potential, which is attributed to properties of corresponding single-particle localized states in the absence of interaction.

\item For quasi-1D ($d\times L$, with $d \ll L$) and 2D ($L\times L$) random systems, the MPS-TDVP data indicate an unbounded growth of $W_c$ in the limit of large $d$ and $L$. The results are in good agreement with analytical predictions \eqref{eq:q1dWc} for the critical disorder $W_c(L,d)$ based on the rare-event avalanche theory. Thus, the MPS results support the prediction of the avalanche theory that the MBL phase is destabilized in 2D in the thermodynamic limit if considered at fixed disorder $W$. At the same time, there is a well-defined MBL transition at a size-dependent disorder $W_c(L,d)$.

\end{enumerate}

The MPS approach is thus a very fruitful avenue for analyzing MBL-type systems. Its key advantage over exact diagonalization is a possibility to study the quantum dynamics in large systems, with $\approx 100$ ``qubits,'' as in current (and presumably near-future) experiments with quantum devices.  The step from $\approx 20$ (as accessible to exact diagonalization) to 100 ``qubits'' is of great importance. However, not unexpectedly, the MPS approach also has its limitations. The main drawback of exact diagonalization---the limitation to relatively small \emph{sizes}---is replaced by a cutoff in time. Longer simulation times can ameliorate this issue somewhat, especially at strong disorder, but eventually the growth of entanglement will become an insurmountable obstacle. This means that conclusions about the critical disorder are drawn based on an assumption that the time scale accessed by the method is sufficient to judge about the long-time limit. As was emphasized, the agreement between $W_c$ obtained from the MPS approach and from the exact diagonalization for relatively small systems clearly supports this assumption. Nevertheless, it is worth emphasizing the following ``disclaimer'' that applies quite generically to investigations of phase transitions: numerical results concerning the phase transition are most useful when they supplement analytical (and, if available, experimental) results; by themselves, they provide no rigorous proof concerning the ultimate phase diagram, position of the critical point, and critical behaviour.

Before closing the paper, we list a few perspective directions for future research.

\begin{enumerate}

\item One important direction concerns a possible optimization of the MPS-type approaches beyond 1D geometry. In Ref.~\cite{Doggen2020a}, a mapping of the 1D structure of an MPS to a ``snake''-like structure was used. It is not obvious a priori that this is the most efficient way, since it does not fully take advantage of the local structure of the lattice in 2D geometry. While it appears that the MPS mapping used in Ref.~\cite{Doggen2020a} is more numerically efficient than a PEPS implementation \cite{Kshetrimayum2019a} that employs a 2D structure into the tensor network, it is conceivable that a still more efficient approach might be devised, perhaps on the basis of tree tensor networks (TTN) \cite{Kloss2020a} or the multiscale entanglement renormalization ansatz (MERA) \cite{Vidal2007a}.

\item  Most of the research of the quantum dynamics within the MPS approach up to now has addressed the imbalance dynamics. Other initial states and other observables---especially those that can be studied in experiment---are also of interest.  In particular, it would be interesting to develop further the investigations of the domain-wall melting started in Ref.~\cite{Hauschild2016a}.

\item Another very intriguing direction is a study of the dynamics of ergodicity avalanches in specially devised settings, see, e.g., a recent experimental work \cite{leonard2020signatures}.

\item A very interesting---and also very challenging---prospect is the investigation of the critical behaviour at the MBL transition (in particular, verification of RG theories predicting the BKT-like scaling near the transition).

\end{enumerate}

\section{Acknowledgments}

We acknowledge collaboration with T. Neupert, F. Schindler, and K. Tikhonov on Ref.~\cite{Doggen2018a}, which was the paper that started our activity on the MPS-based studies of the MBL problem. In course of these studies, we enjoyed useful discussions with many colleagues, including F. Alet, Y. Bar Lev, I. Bloch, F. Evers, M. H. Fischer, S. Gopalakrishnan,  S. Goto, M. Heyl, C. Karrasch, M. Knap, N. Laflorencie, D. Luitz, S. R. Manmana, M. M\"uller, R. M. Nandkishore, A.~Polkovnikov, S. Rex, A. Scardicchio, B. I. Shklovskii, K. Tikhonov, T. Wahl, J. Zakrzewski, and M. \v{Z}nidari\v{c}. The authors acknowledge support by the state of Baden-W\"urttemberg through bwHPC.

\bibliography{ref}

\end{document}